\documentclass[12pt]{article}
\usepackage{graphicx,color}
\usepackage{amsmath}
\usepackage{amssymb}
\usepackage{xspace}
\usepackage{mathrsfs}
\usepackage{subfigure}
\usepackage{overpic}
\usepackage{hyperref}
\definecolor{webgreen}{rgb}{0,.5,0}
\definecolor{webbrown}{rgb}{.6,0,0}
\definecolor{RoyalBlue}{rgb}{0,0,.5}
\usepackage{lineno}
\hypersetup{%
     colorlinks=false, linktocpage=false, pdfborder={0 0 0}, pdfstartpage=3,
     pdfstartview=FitV
}

\newcommand\pubnumber{}
\newcommand\pubdate{\today}

\def\Title#1{\begin{center} {\Large #1 } \end{center}}
\def\Author#1{\begin{center}{ \sc #1} \end{center}}
\def\Address#1{\begin{center}{ \it #1} \end{center}}

\newcommand\pubblock{\rightline{\begin{tabular}{l} \pubnumber\\
         \pubdate  \end{tabular}}}
\newenvironment{Abstract}{\begin{quotation}  }{\end{quotation}}
\newenvironment{Presented}{\begin{quotation} \begin{center} 
             PRESENTED AT\end{center}\bigskip 
      \begin{center}\begin{large}}{\end{large}\end{center} \end{quotation}}

\def\beq{\begin{equation}}
\def\eeq#1{\label{#1}\end{equation}}
\def\eeqn{\end{equation}}

\def\beqa{\begin{eqnarray}}
\def\eeqa#1{\label{#1}\end{eqnarray}}
\def\eeqan{\end{eqnarray}}

\let\bar=\overbar

\def\Dslash{\not{\hbox{\kern-4pt $D$}}}
\def\dslash{\not{\hbox{\kern-2pt $\del$}}}

\def\msb{{\bar{\ssstyle M \kern -1pt S}}}

\newcommand{\dmumu}{\ensuremath{D^0 \to \mu^{+} \mu^{-}}\xspace}
\newcommand{\dee}{\ensuremath{D^0 \to e^{+} e^{-}}\xspace}
\newcommand{\dellell}{\ensuremath{D^0 \to \ell^{+} \ell^{-}}\xspace}
\newcommand{\demu}{\ensuremath{D^0 \to e^{\pm} \mu^{\mp}}\xspace}
\newcommand{\dgammagamma}{\ensuremath{D^0 \to \gamma \gamma}\xspace}
\newcommand{\dpimumu}{\ensuremath{D^+ \to \pi^+ \mu^{+} \mu^{-}}\xspace}

\def\dstdpipi{\ensuremath{D^{\ast +}\to D^0 (\to \pi^{+} \pi^{-}) \pi^+}\xspace}

\def\dstdpi{\ensuremath{D^{\ast +}\to D^0 \pi^+}\xspace}
\def\dm{\ensuremath{\Delta M}\xspace}

\newcommand{\dpipi}{\ensuremath{D^0 \to \pi^{+} \pi^{-}}\xspace}
\newcommand{\dkpi}{$D^0 \to K^{-} \pi^{+}$\xspace}
\def\ctoull{\ensuremath{c\to u \ell^+ \ell^-}\xspace}
\def\xzeroctohll{\ensuremath{X^0_c\to h^0 \ell^+ \ell^-}\xspace}
\def\xplusctohll{\ensuremath{X^{\pm}_c\to h^{\pm} \ell^+ \ell^-}\xspace}
\def\babar{\textsc{BaBar}\xspace}
\def\dzero{$\rm{D\varnothing}$\xspace}
\def\fb{\ensuremath{\rm{fb}^{-1}}\xspace}
\def\pb{\ensuremath{\rm{pb}^{-1}}\xspace}
\def\gevcc{\ensuremath{\rm{GeV/c^2}}\xspace}
\def\mevcc{\ensuremath{\rm{MeV/c^2}}\xspace}

\begin{document}
\begin{titlepage}
\pubblock

\vfill
\Title{Rare charm decays: an experimental review}
\vfill
\Author{ Francesco Dettori} 
\Address{NIKHEF-VU 
Science Park 105
1098 XG Amsterdam 
Netherlands  (fdettori@nikhef.nl)}
\vfill
\begin{Abstract}
Indirect searches, and in particular rare decays,  have proven to be a fruitful
field to search for New Physics beyond the Standard Model. While the
down-quark sector (B and K) have been studied in detail, less attention was
devoted to charm decays due to the smaller expected values and
higher theoretical uncertainties of their observables. 
Recently a renewed interest is growing in rare charm searches.
In this article we review the current experimental status of searches for rare
decays in charmed hadrons. While the Standard Model rates are yet to be
reached, current experimental limits are already putting constraints on New
Physics models. 
\end{Abstract}
\vfill
\begin{Presented}
Charm 2012\\ The 5th International Workshop on Charm Physics\\
14-17 May 2012, Honolulu, Hawai'i 
\end{Presented}
\vfill
\end{titlepage}
\def\thefootnote{\fnsymbol{footnote}}
\setcounter{footnote}{0}

\section{Introduction}

The Standard Model (SM) of particle physics is an exceptionally precise theory
and up to now is able to describe all the available experimental information,
with really few exceptions. Therefore deviations from this theory belong either
to non-explored energy regimes or to extremely small effects.
While direct searches at current accelerators have not yet led to New Physics
(NP) discoveries, hints of new effects might hide in rare processes. 

Due to the flavour structure of the SM, flavour changing neutral currents
(FCNC) are only allowed at loop level, giving the possibility to explore
effects where the dominant tree level is forbidden. Furthermore they give
access, through the virtual loops, to higher energy ranges than those
accessible to direct searches. 
FCNC and rare decays in general have been extensively studied in the $down$-type
quark sector, \emph{i.e.}
in B and K meson decays, and have proven to constrain tightly New Physics
possibilities. Less attention was devoted in the past to the $charm$
sector, mainly due to the fact that SM values of FCNC for charmed hadrons are
very small. In fact, the absence of a heavy down-type quark which would
decouple the short distance effects (as the $top$ quark does for the $B$ and
$K$ decays) implies that the Glashow-Iliopoulos-Maiani (GIM) mechanism
\cite{Glashow:1970gm} is extremely effective for charmed hadrons. 
On one side this leads to the advantage that NP effects can be orders of
magnitude larger than the SM ones; however it also implies that SM predictions
can be dominated by long distance contributions due to the propagation of light
quarks intermediate states. Unfortunately long distance contributions, being
non perturbative, carry large theoretical uncertainties. 

Many New Physics scenarios can enhance these processes, like Supersymmetry
(with or without $R$-parity violation)~\cite{Burdman:2001tf}, Littlest Higgs
model~\cite{Lee:2004me,Fajfer:2005ke} and Leptoquarks~\cite{Fajfer:2008tm}.
Furthermore some models lead to enhancements only in the $up$ sector,
and therefore can only be probed in charm decays.

As the experimental results improve a renewed interest in the
charm sector has grown, also due to the recent measurements of CP violation in
D decays by the LHCb~\cite{Aaij:2011in} and CDF
experiments~\cite{Aaltonen:2011se}, stimulating the theoretical work. 

We will review in the following the current status of experimental searches for
rare charm decays. In particular in \S\ref{sec:ctoull} we will treat the FCNC
\ctoull ~\footnote{Here and in the rest of this article we always imply charge
conjugation unless otherwise stated.}
 transitions for charged  (\S\ref{sec:charged_fcnc}) and neutral charm
hadrons  (\S\ref{sec:neutral_fcnc}).
Radiative processes will not be treated in the following as
the high theoretical uncertainties make them less
interesting~\cite{Burdman:2001tf}, though recently it has been proposed that
their study could shed light on charm CP violation~\cite{Isidori:2012yx}. 
In sections \ref{sec:dgammagamma} and
\ref{sec:dellell} we will describe the di-photon and di-lepton $D^0$
decays. Finally a brief overview of the current searches for SM forbidden
decays will be given in section \ref{sec:forbidden}. 

As far as the future prospects are concerned, we will not discuss them in this
article as they are subject of another report at this same conference.

\section{\ctoull transitions} 
\label{sec:ctoull}
We start from the \ctoull transitions as they are the most probable to be
observed at currently running experiments. Short distance contributions for
transitions of the type \ctoull are heavily
suppressed by the GIM mechanism. In particular, the inclusive branching
fractions calculated for the
mesonic modes to electrons are \cite{Burdman:2001tf}:
$${\cal B}_{D^+ \to X^+_u e^+ e^-} \simeq 2 \cdot 10^{-8}  \qquad\qquad
{\cal B}_{D^0 \to X^0_u e^+ e^-} \simeq 8 \cdot 10^{-9}  ,$$ 
which would be out of present experimental reach. 
However, as already said, short distance contributions are shaded by the long
distance ones. 
Long distance (LD) processes can be represented in the form $D\to X V \to X \ell
\ell$ where $V = \phi, \rho, \omega$ is the intermediate resonance. 
Typical exclusive branching fractions of the resonant processes are at the level
of $\mathcal{O}(10^{-6})$~\cite{Burdman:2001tf, Singer:1996it}.
As an example the branching fraction for $D^+ \to \pi^+ e^+ e^-$ is
$\mathcal{B}\simeq 2 \times 10^{-6}$ dominated by the $\pi^+ \phi$ intermediate
state. 

However long distance processes, being non-perturbative in nature, carry also
large theoretical uncertainties so that they make more difficult the comparison
between SM and New Physics predictions and between these and the experimental
results. 

Different New Physics scenarios can enhance \ctoull transitions. 
Already the Minimal Supersymmetric Standard Model (MSSM) with R-parity
conservation could give contributions to \ctoull processes.
 These would come from one-loop diagrams
with gluino and squarks in the loop \cite{Burdman:2001tf,Burdman:2003rs} and
would increase the decay rate at low di-lepton invariant mass. However when
including bounds from $D^0 -\bar D^0$ oscillations these enhancements are
reduced. Furthermore when considering the hadronic decay
$D\to X \ell^+ \ell^-$ the rate is decreased by a factor $m^2_{\ell \ell}$ when
$X$ is a pseudoscalar meson and hidden by long distance contributions when $X$
is a vector meson  \cite{Fajfer:2007dy}.

\begin{figure}
\subfigure[]{\includegraphics[width = 0.5 \textwidth]{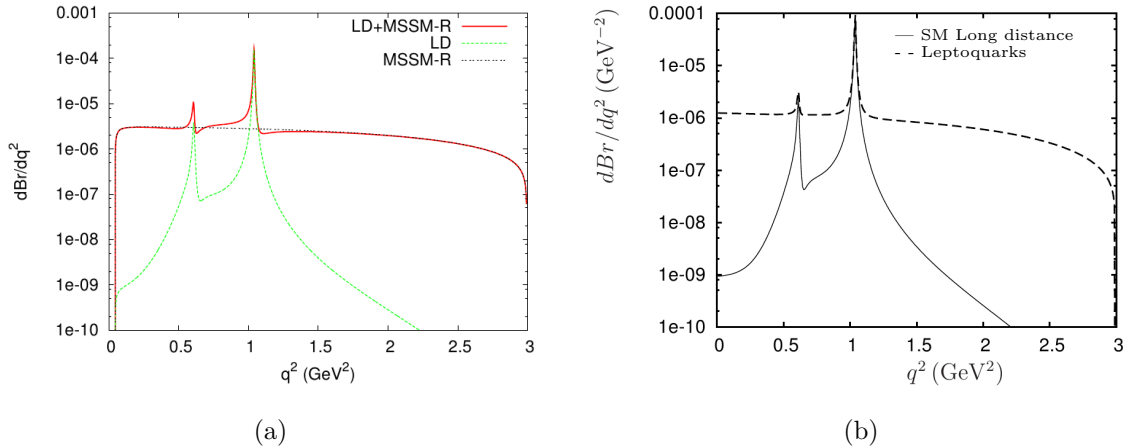}}
\subfigure[]{
\begin{overpic}[width = 0.5 \textwidth ]{Dpimumu_leptoquark_2.pdf}
 \put(55,60){{\tiny - - Leptoquarks}}
 \put(55,63){{\tiny --- SM Long distance}}
\end{overpic}
}
\caption{Differential rate as a function of invariant mass squared of the two
muons in the \dpimumu decay for the SM long distance (LD) prediction, and
compared with (a) MSSM$\displaystyle{\not} R_p$~\cite{Fajfer:2007dy}
and (b) Leptoquarks new physics model. The Leptoquark contribution is saturated
to the
experimental limit on the total branching fraction~\cite{Abazov:2007aj}.
}\label{fig:NP_pimumu} 
\end{figure}

On the other hand the MSSM with R-parity violation ($\displaystyle{\not} R_p$) 
is not so constrained and can give large contributions via a tree-level
exchange of squarks. Enhancement of a factor $\sim3$ are possible, for example,
in the \dpimumu decay \cite{Fajfer:2007dy}. In particular, as shown in
Fig.~\ref{fig:NP_pimumu}(a), at low and high di-muon
invariant mass (or $q^2 = m^2_{\ell \ell}$), \emph{i.e.} far from the
resonances, there
is a possible high sensitivity to the MSSM$\displaystyle{\not} R_p$. 

Similar increases in the branching fraction can appear also in the $D\to \rho
\mu^+ \mu^-$ channel; in this channel in addition to the branching fraction also
angular observables, such as the forward-backward asymmetry for the leptons are
thought to be sensitive to NP effect, even if only reachable at high
yields \cite{Burdman:2001tf,Fajfer:2007dy}. 

Other possible contributions to \ctoull can come from new physics leptoquarks. 
In particular it has been shown that the \dpimumu decay is sensitive also to
 couplings which cannot be limited by constraints on the
branching fraction of \dmumu \cite{Fajfer:2008tm}. In
Fig.~\ref{fig:NP_pimumu}(b)
the differential branching fraction of \dpimumu is shown in the hypothesis of
saturating its experimental limit with Leptoquarks coupling, showing that large
room for a distinction between this and the LD SM contribution is still present
at large and low $q^2$.

In the next two sections we will review the experimental searches for \ctoull
transitions divided for simplicity into charged and neutral hadron decays. 

 \subsection{Searches for \xplusctohll}
\label{sec:charged_fcnc}

\begin{figure}
\subfigure{\begin{overpic}[width = 0.8 \textwidth ]{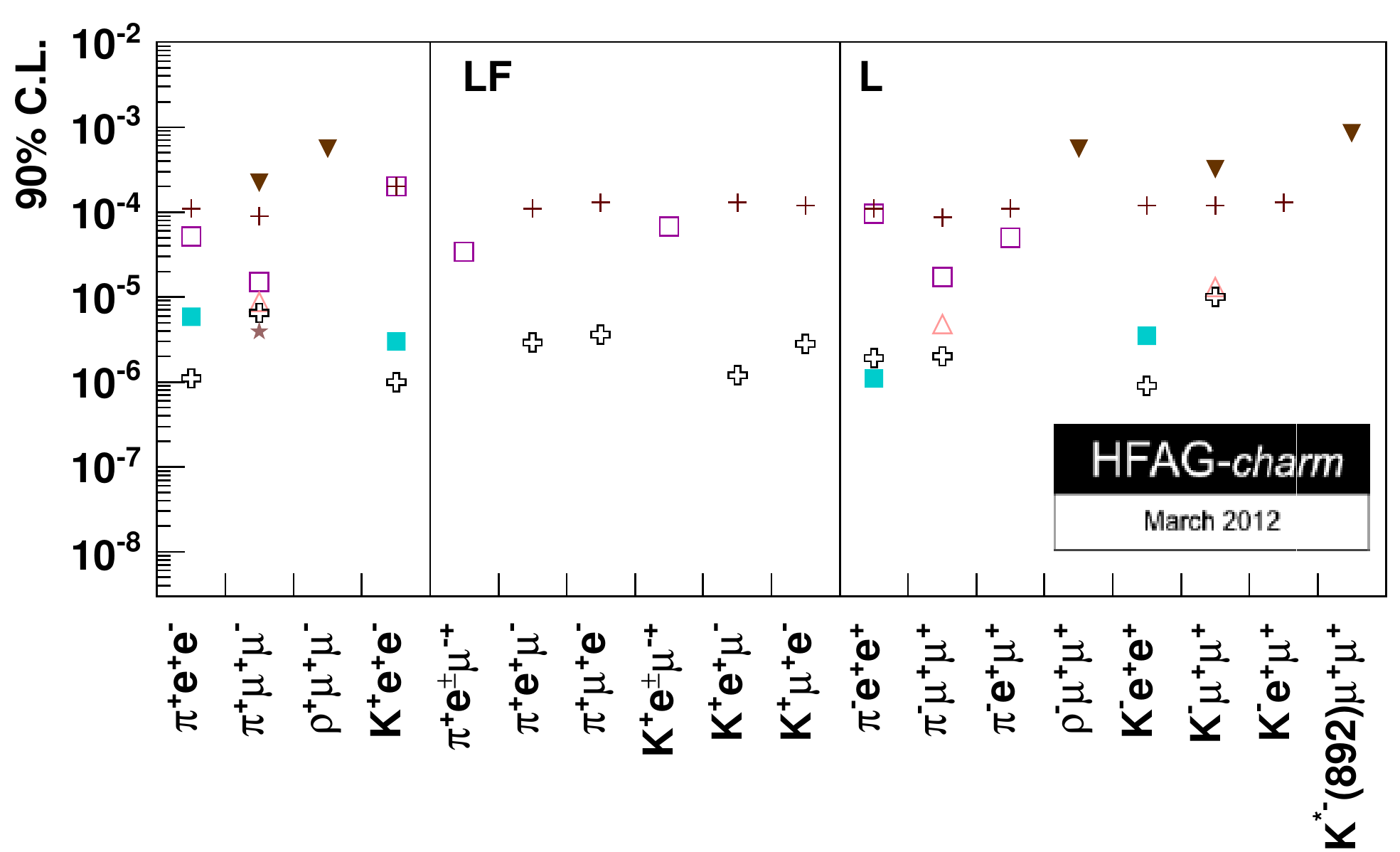}
 \put(20, 25){$D^+$ decays}
\end{overpic}}
\subfigure{\begin{overpic}[width = 0.8 \textwidth ]{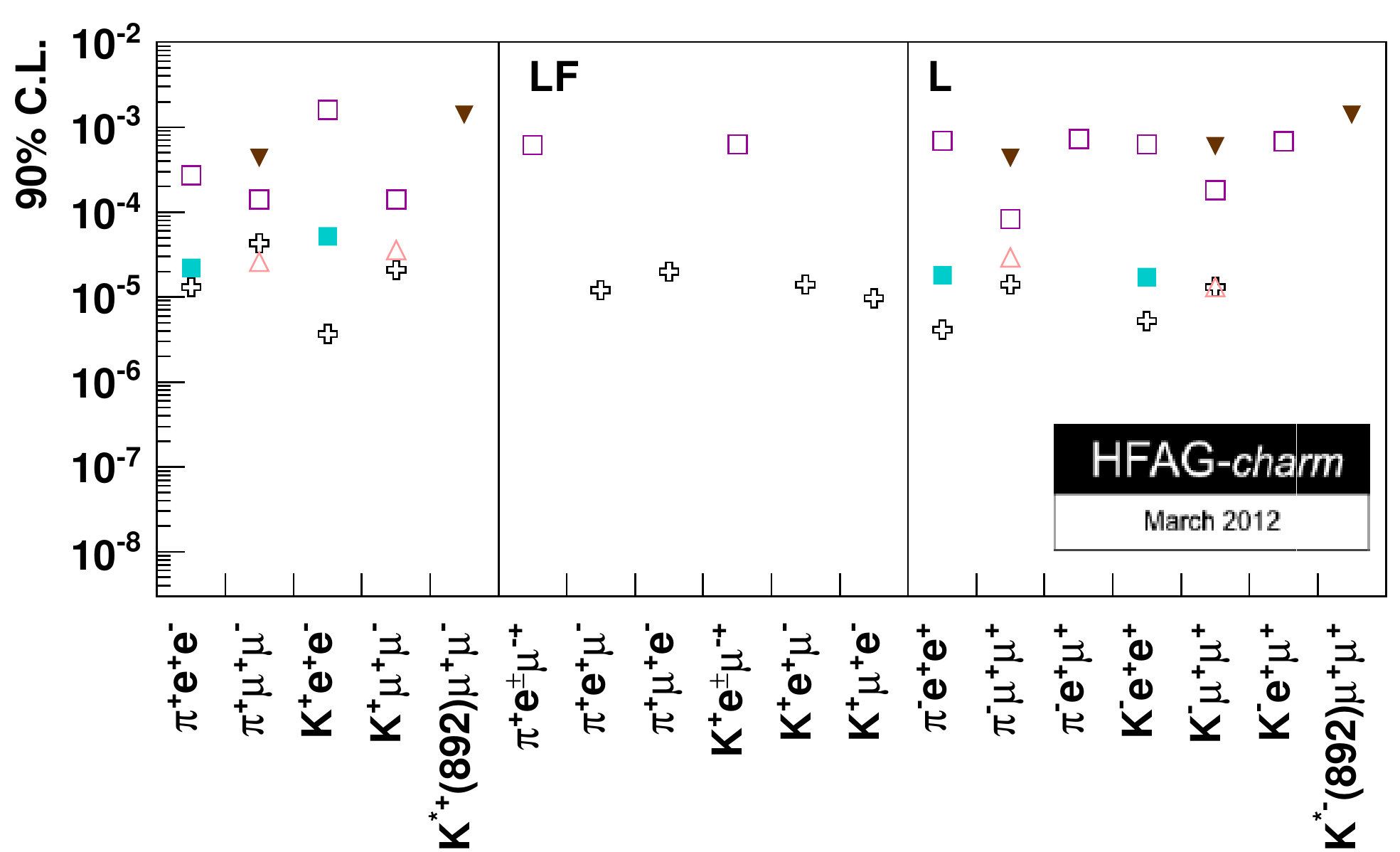}
 \put(20, 25){$D^+_{s}$ decays}
\end{overpic}}
\subfigure{\includegraphics[width = 4 cm ]{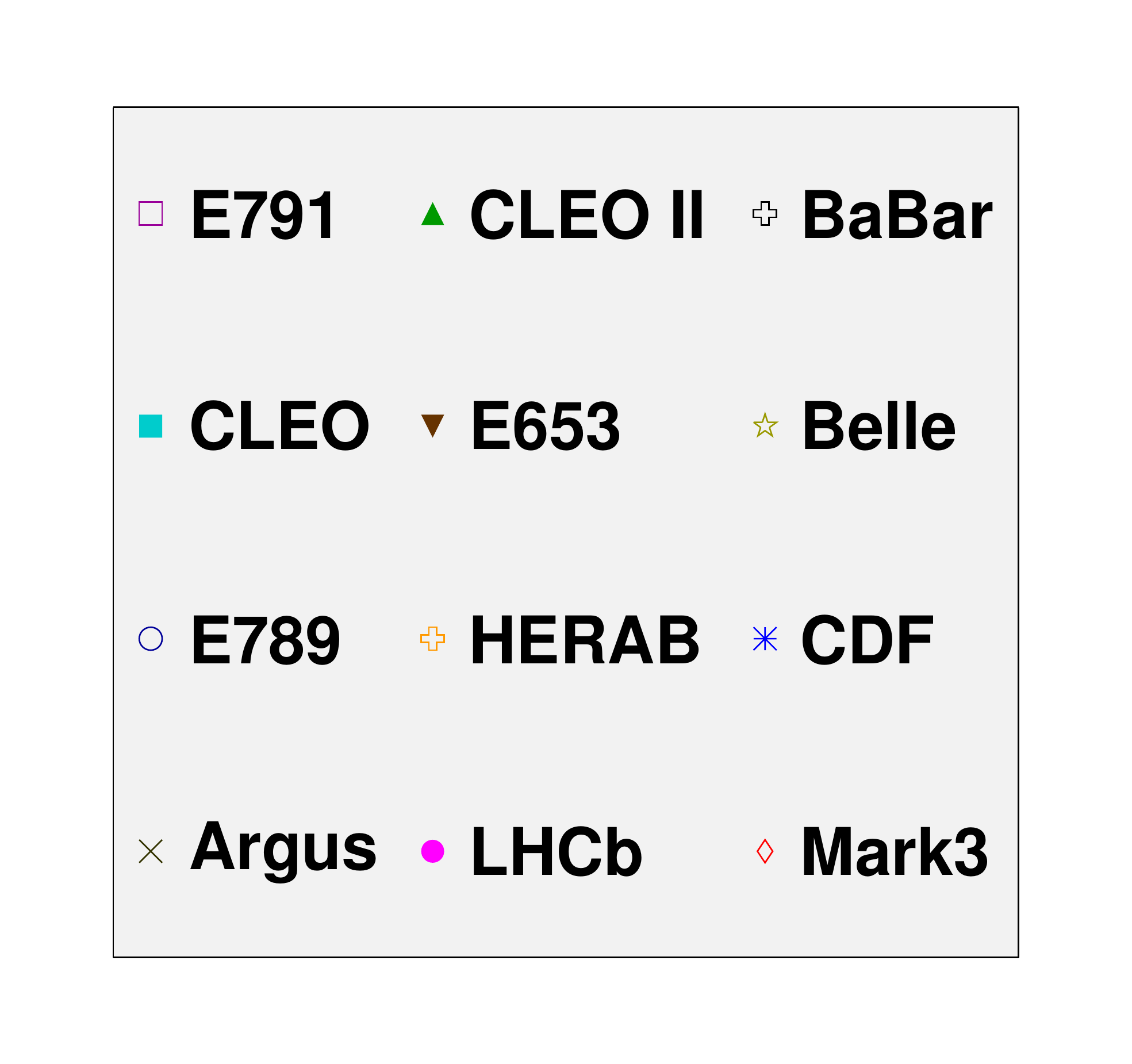}}
\subfigure{\begin{overpic}[width = 0.35 \textwidth]{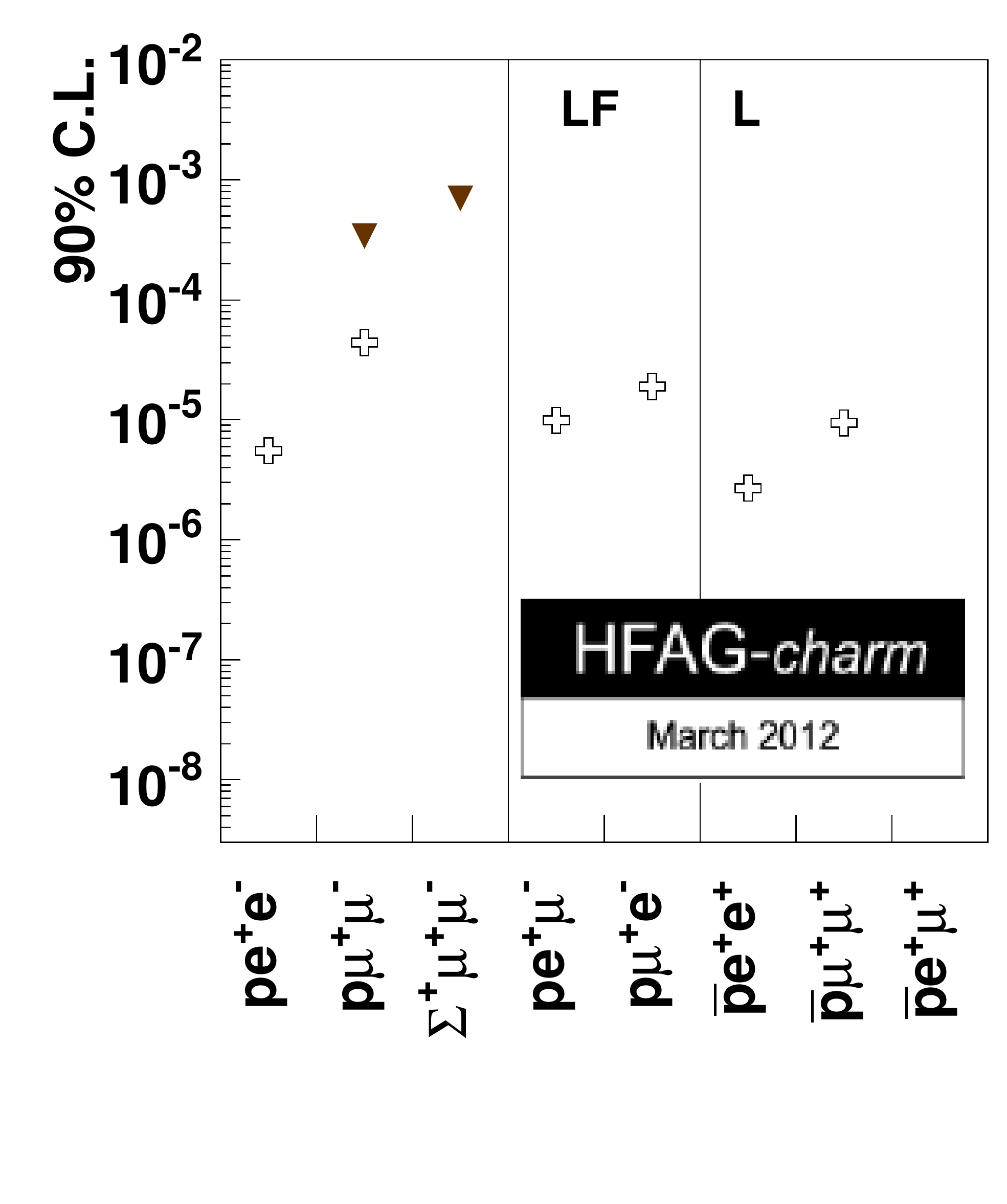} 
 \put(45, 80){$\Lambda_{c}^+$ decays}
\end{overpic}}
\caption{Summary of the upper limits on charged charm hadron rare decays  
compiled by the HFAG group. \cite{Amhis:2012bh}}\label{fig:hfag_charged}
\end{figure}

The current limits on rare decays of charged charm hadrons are nicely
summarized by the Heavy Flavour Averaging Group (HFAG)~\cite{Amhis:2012bh} plots
shown in Fig.~\ref{fig:hfag_charged}.

None of the listed channels have been observed up to now and the limits on the
decays branching fractions are at the level of $10^{-6}$ or higher. 
With few exceptions the best limit for all of these decay modes come from a
comprehensive analysis of the \babar  experiment, described
in Ref.~\cite{Lees:2011hb}, which we review in the following. 
  
\subsubsection{Search for \xplusctohll at \babar}
\label{sec:charged_fcnc_babar}

A search for rare decays in the form \xplusctohll was conducted by the \babar
collaboration \cite{Lees:2011hb} exploiting data collected at the PEP-II at
SLAC. A dataset of 347~\fb of $e^+e^-$ collisions at the $\Upsilon(4S)$ and of
37 \fb just below it was used. We will not describe the \babar detector
and the reader should refer to Ref.~\cite{Aubert:2001tu}. 

Pairs of tracks identified as leptons and a track identified as a charged hadron
(pion, kaon or proton) were used to build charm hadron candidates. 
The candidate momentum in the $e^+e^-$ center of mass ($p^{\ast}$) was required
to be larger than 2.5 \gevcc to remove a large combinatorial
background in those regimes. 
QED background was rejected requiring a multiplicity larger than 5. The 
residual combinatorial background, mainly from charm and beauty semi-leptonic
decays, was reduced with cuts on the $\chi^2$ of the vertex fit and on the
distance of closest approach between the leptons.
Finally the $\phi$ region was rejected from the invariant mass of the
di-lepton in order to exclude the resonant (and long distance) part of the
decay. 
After this first selection a likelihood ratio was used to further discriminate
the signal from combinatorial background. The likelihood was built from
information on the $p^{\ast}$ of charmed candidate, total energy in the event
and flight length significance. The PDFs for the signal were built using Monte
Carlo (MC) simulations while for the background data events in the
invariant mass sidebands were used. 
The cut value on the likelihood was optimized independently for each
decay mode in order to provide the lowest expected upper limit. 

Extended, unbinned maximum-likelihood fits to the invariant mass
distributions were used in order to extract the number of signal events. The
signal pdf was parametrized as a Crystal Ball function while the
combinatorial background as a first order polynomial. For some channels an
additional component to take into account background from mis-identified non
leptonic charm decays was included. 
In Fig.~\ref{fig:fcnc_babar} the invariant mass distribution of some of the
studied channels is shown together with the fit. 

The signal yields were normalized to known hadronic charm decays in order to
compute the branching fractions: $D^+_{(s)}$ mesons channels were normalized to
the $D^+_{(s)} \to \phi(\to KK) \pi^+$ channel while $\Lambda_c^+\to p K^-
\pi^+$ decay was used for the $\Lambda_c^+$ decays. 

No significant signal was observed in any of the channels. A 2.6$\sigma$
fluctuation was seen in the $\Lambda^+_c \to p \mu^+\mu^-$ decay which however
has a 25\% probability to occurr when looking at 35 different channels.
Therefore upper limits were put on the decay branching fractions using a
Bayesian approach with
flat prior for the event yield in the physical region. 
The obtained upper limits on the branching fractions at 90\% CL are between
$1\cdot10^{-6}$  and $44\cdot 10^{-6}$ depending on the decay  mode  and are the
best limits up to now for most of the channels.

\begin{figure}
\subfigure{\includegraphics[width = 0.5 \textwidth]{babar_Dpill.pdf}}
\subfigure{\includegraphics[width = 0.5 \textwidth]{babar_DKll.pdf}}
\subfigure{\includegraphics[width = 0.5 \textwidth]{babar_Lcpll.pdf}}
\caption{Invariant mass distributions for \xplusctohll
candidates in the analysis performed by \babar
\cite{Lees:2011hb}.}\label{fig:fcnc_babar}
\end{figure}

\subsubsection{\dpimumu at \dzero}
\label{sec:charged_fcnc_dzero}

One of the channels in which the best limit does not come from \babar is the
\dpimumu decay were a strongest limit was put by the \dzero collaboration
\cite{Abazov:2007aj}. 
This search was conducted with 1.3 \fb of $p\bar p$ collisions at
$\sqrt{s} = 1.96$~TeV collected at Tevatron with the \dzero detector. 

\begin{figure}
 \subfigure{\begin{overpic}[width =0.5
\textwidth]{D0_Dpimumu_phi.pdf}
  \put(20,20){$D\to \phi(\to \mu \mu) \pi$}
 \end{overpic}}
 \subfigure{\includegraphics[width = 0.5
\textwidth]{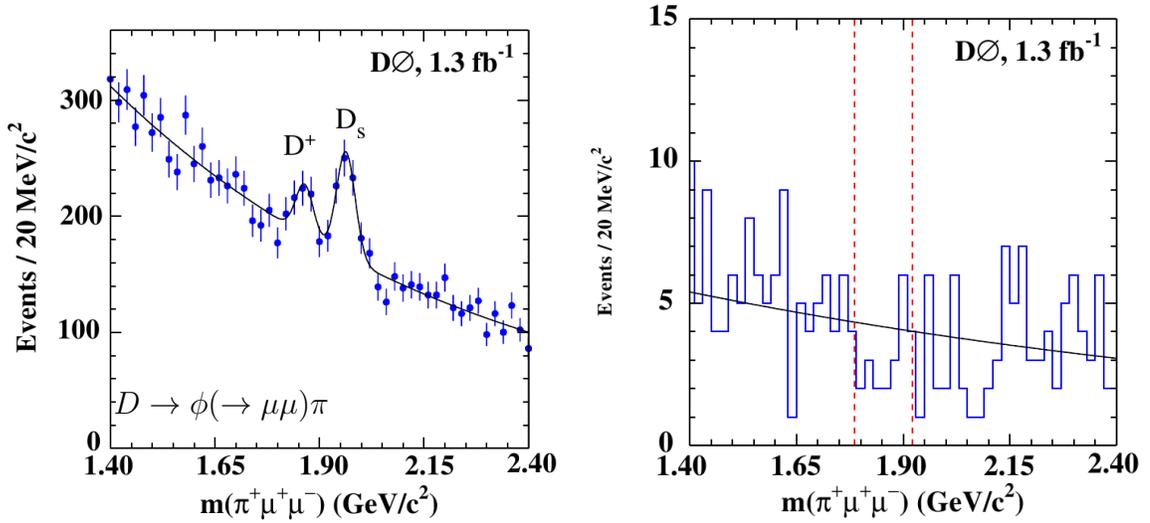}}
\caption{Invariant mass distribution of $D^+_{(s)}\to
\pi^+ \mu^+ \mu^-$candidates for events with the di-muon (a)
in the $\phi$ mass region and (b) outside of it \cite{Abazov:2007aj}.
}\label{fig:dpimumu_d0}
\end{figure}

Tracks identified as muons were used as starting point after a cut on the
transverse momentum (2 GeV/c) and on the momentum (3 GeV/c). Two oppositely
charged muons were combined requiring to form a well reconstructed vertex and
to have an invariant mass lower than 2 \gevcc. 
The $\phi$ invariant mass region of the di-muon system was excluded from the
search of the non-resonant \dpimumu system and used to study the $D^+ \to \phi
\pi^+$ decay, with $\phi\to \mu^+ \mu^-$. 
The di-muon system was combined with an additional track passing requirements on
its impact parameter significance and on its transverse momentum, and the three
body invariant mass was required to be in the 1.4 - 2.4 \gevcc range. 
Further selection criteria were applied based on the significance of the
transverse flight distance of the $D^+$ candidate and on the collinearity angle,
\emph{i.e.} the angle between the $D^+$ momentum and the line between primary
and secondary vertices.
After this selection the $\pi^+ \mu^+ \mu^-$ invariant mass for events in the
$\phi$ di-muon mass region is shown in Fig.~\ref{fig:dpimumu_d0}(a), where the
two signals of $D^+$ and $D^+_s \to \phi \pi^+$ can be seen. The $D^+ \to \phi
\pi^+ \to \mu^+\mu^-\pi^+$ branching fraction was measured to be 
$(1.8 \pm 0.5 (\rm{stat}) \pm 0.6(\rm{syst}))\times 10^{-6}$. 
Before considering the non-resonant part of the di-muon invariant mass, tighter
cuts were applied based on the already mentioned variables and on the pion
transverse momentum and the isolation of the $D^+$: $\mathcal{I} =p_D/\sum
p_{cone}$ where the sum of the momenta is extended to a cone around the
$D^+$ candidate. The criteria were optimized in order to obtain the best upper
limit on the branching fraction. 
The invariant mass distribution of the remaining events is
shown Fig.~\ref{fig:dpimumu_d0}(b). No signal above the combinatorial
background was observed and the following upper limit on the branching fraction
was estimated: $\mathcal{B}(\dpimumu)< 3.9 \times 10^{-6}$ at 90\% CL.

While this limit is still more than 2 orders of magnitude from the Standard
Model expected rate it constraints already the parameter space of 
R-parity violating SUSY~\cite{Burdman:2001tf}; on the other hand the
little Higgs models are an order of magnitude below this level and
therefore are still not constrained by this measurement~\cite{Fajfer:2005ke}.

 \subsection{Searches for \xzeroctohll}
\label{sec:neutral_fcnc}

\begin{figure}
 \begin{overpic}[width = 0.8 \textwidth ]{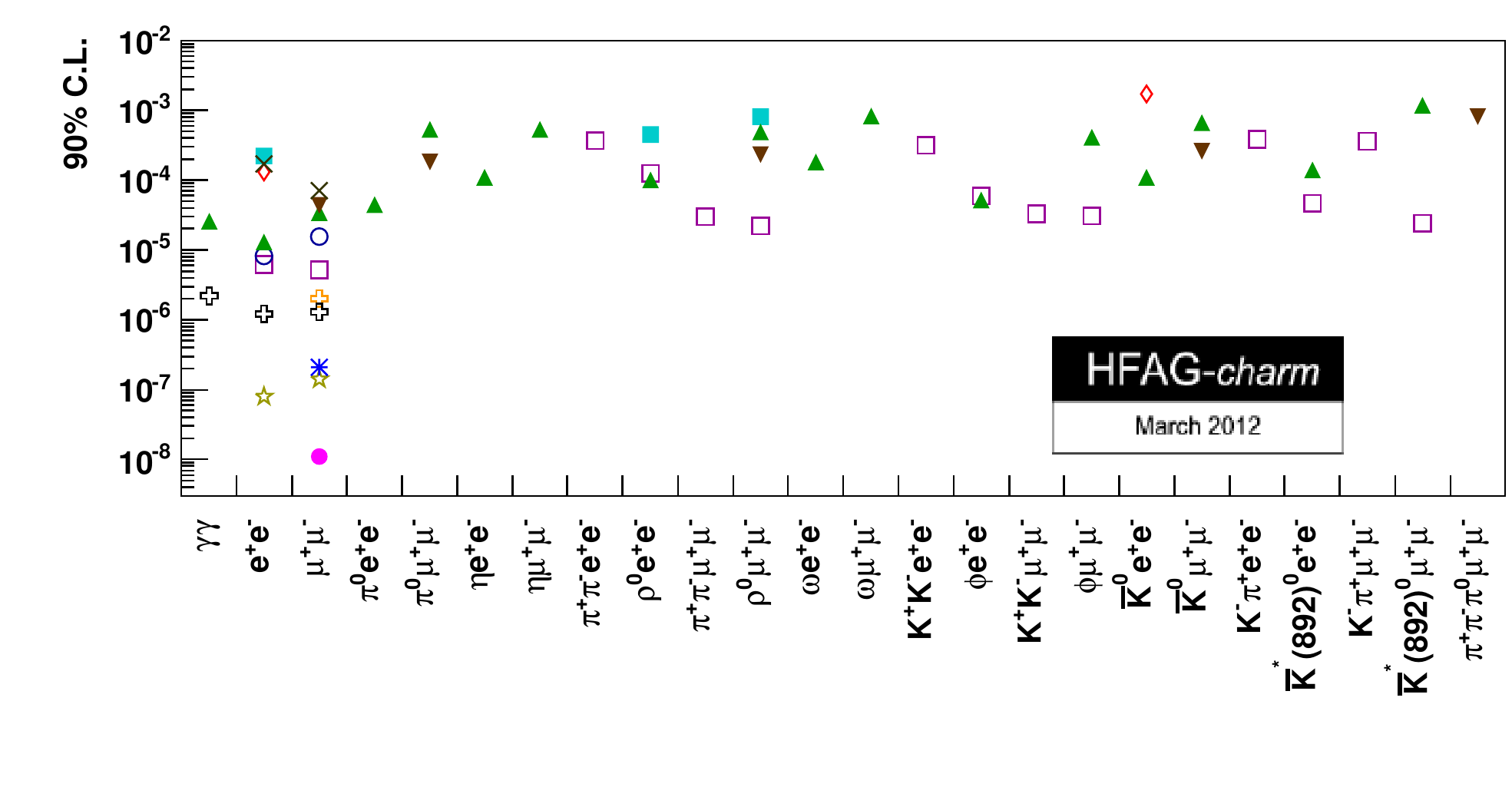}
\put(25, 30){$D^0$ decays}
 \put(100, 30){\includegraphics[width = 2.5 cm ]{D0_leg.pdf}}
\end{overpic}
\caption{Summary of upper limits on branching fraction of FCNC $D^0$ decays
\cite{Amhis:2012bh}.}\label{fig:hfag_dzero}
\end{figure}

Results on rare decays from $D^0$ decays are all at least 10-years old, with
the exception of di-lepton and di-photon decays which will be treated
separately. 
A summary of the upper limits on the various decay modes is shown in
Fig.~\ref{fig:hfag_dzero} from the HFAG group~\cite{Amhis:2012bh}. With few
exceptions, all the limits on $D^0 \to X
\ell \ell$ decays come from two analyses by the E791 and CLEOII
collaborations which will be discussed in the following. 

\subsubsection{Search for $D^0 \to X \ell \ell$ decays at E791 }
\label{sec:neutral_fcnc_e791}

\begin{figure}
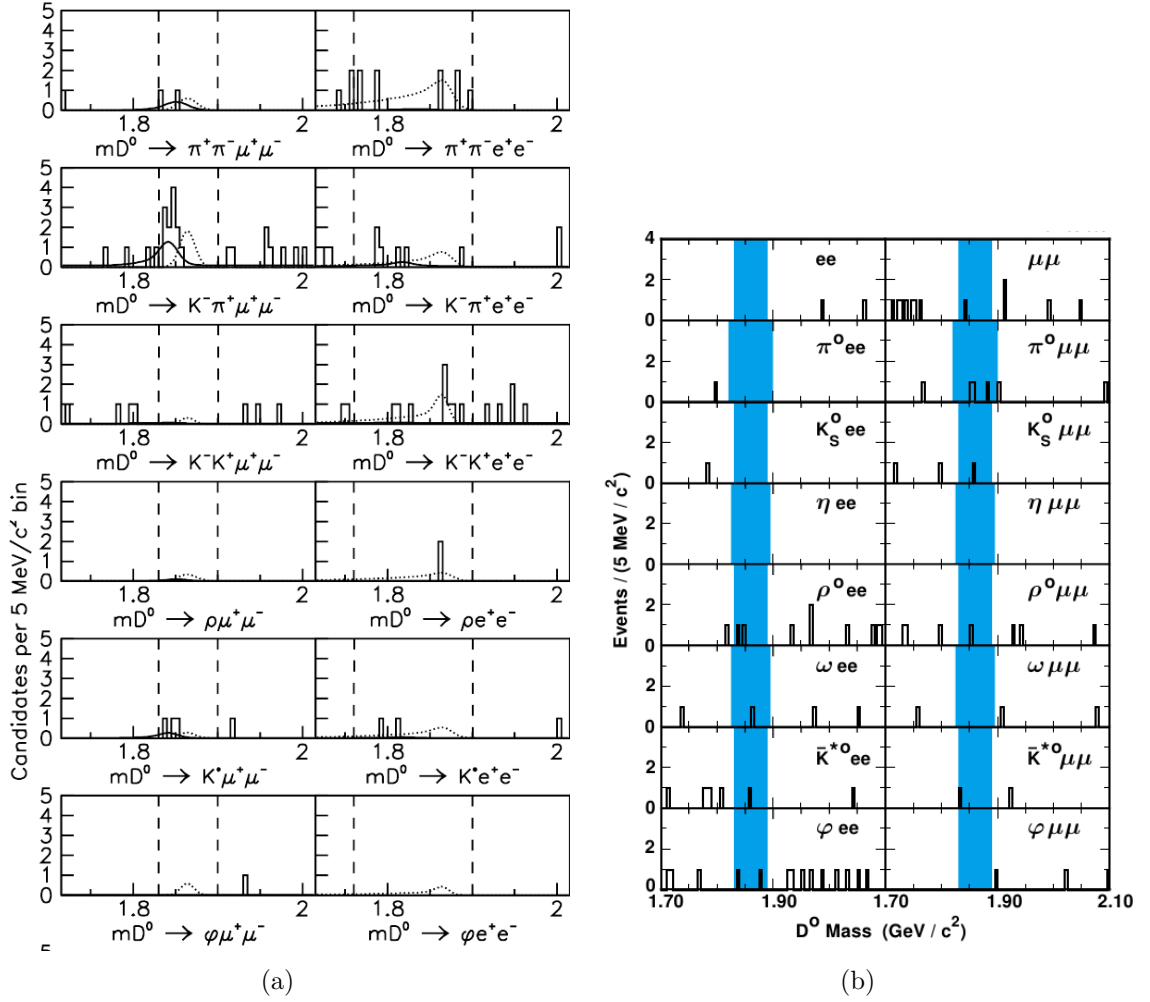

\subfigure[]{\includegraphics[width = 0.5 \textwidth]{E791_d0hmumu.pdf}}
\subfigure[]{\includegraphics[width = 0.5 \textwidth ]{CLEOII_results.pdf}}
 \caption{Invariant mass distributions for (a) the search of signals
in the form \mbox{$D^0 \to X \ell \ell$} performed by the E791
collaboration~\cite{Aitala:2000kk} and (b) candidates in the form $D^0 \to X
\ell \ell$ in the search performed by the CLEO
collaboration~\cite{Freyberger:1996it}.
}\label{fig:e791_cleo_fcnc}
\end{figure}

The E791 collaboration reported in 2001 a search for decays in the forms 
$D^0 \to V \ell^+ \ell^-$, with $V = \rho^0, \bar K^{\ast 0}, \phi$ and $D^0 \to
h h \ell^+ \ell^-$, with $h=\pi, K$ \cite{Aitala:2000kk} based on $2\times
10^{10}$ events of a 400 GeV/c $\pi^-$ beam on fixed target. 
Events with secondary vertex well separated from the primary one and from the
detector material were selected. 
A blind analysis was optimized exploiting MC simulations of the signal channels
and using background from the invariant mass sidebands. 
All the channels were normalized to topologically similar hadronic decays to
which the very same selection, with the exception of particle
identification, was applied. MC simulations were used in order to calculate the
signal and normalization selection efficiencies, while the muon and electron
identification efficiencies were measured on data. 
After unblinding the signal invariant mass regions no significant signal was
found in any of the channels, as can be seen from the invariant mass
distributions shown in Fig.~\ref{fig:e791_cleo_fcnc}(a). After subtracting the
combinatorial background, estimated from the sidebands, and the
mis-identification
background, upper limits on the number of signal events were calculated with the
Feldman-Cousins method and translated into branching fractions with the
normalisation channels.

The upper limits on the branching fractions vary between 3.0 and 55.3 $\times
10^{-5}$ at 90 \% CL depending on the channel, most of which are the best
results to date.

\subsubsection{Search for $D^0 \to X \ell \ell$ decays at CLEOII }
\label{sec:neutral_fcnc_cleoII}

A similar study was conducted by the CLEO collaboration with the CLEOII
detector at CESR \cite{Freyberger:1996it}. 
A dataset of 3.87 \fb of $e^+e^-$ collisions at the $\Upsilon(4S)$ and just
below it was used to search for events in the form  $D^0 \to X \ell \ell$ 
where $X = \pi^0, K^0_S, \eta, \rho^0, \omega, K^{\ast 0}, \phi$.
In order to build the signal candidates, charged tracks were required to come
from the primary interaction, with the exception of the ones for the $K^0_S$.
Particle identification of pions and kaons was based on $dE/dx$ and
time-of-flight information, while calorimeter and muon chambers information was
additionally exploited for the lepton candidates. Electrons coming from photon
and $\pi^0$ decays were rejected. The various employed mesons were reconstructed
in the following final states: $K^0_S\to \pi^+ \pi^-$, 
$\rho^0 \to \pi^+ \pi^-$, $\omega \to \pi^+ \pi^- \pi^0$, $\bar K^{\ast 0} \to
K^- \pi^+$, $\phi \to K^+K^-$, and $\pi^0$ and $\eta$ into $\gamma \gamma$. 
The so constructed $D^0$ candidate was required to come from a $D^{\ast +} \to
D^0 \pi^+$ decay, in order to reduce the combinatorial background, imposing a
cut on the $D^{\ast +}-D^0$ mass difference ($\Delta M$) at 2.0 \mevcc from the
nominal value. 
Cuts on the scaled momentum of the $D^\ast$ ($p_{D^\ast}/\sqrt{E^2_{beam} -
M^2_{D^{\ast}}}$) and on the angle between the daughter tracks and the $D^0$
momentum were also applied. 

After this selection the invariant mass distributions of the studied decay
modes are shown in Fig.~\ref{fig:e791_cleo_fcnc}(b). No signals were observed
over the
small expected combinatorial background and a 90\% poissonian upper limit was
put on the number of signal events.  
This limit was translated in a branching fraction by normalizing to the total
number of produced $D^{0}$, measured in data in the $K^-\pi^+$ final state, and
correcting for the efficiency of each channel, obtained from simulations. 
The upper limits on the branching fractions of the various modes are in the
range between few in $10^{-5}$ to $10^{-4}$ depending on the channels and are
the best limits to date for most of them.

\section{\dgammagamma}
\label{sec:dgammagamma}

The Standard Model prediction for the short-distance contribution to
the \mbox{\dgammagamma} decay leads to a branching fraction of $3 \times
10^{-11}$,
which however is completely shaded by the long distance
contributions~\cite{Burdman:2003rs}. Calculations done for the Vector Meson
Dominance mechanism give a long distance contribution to the branching
fraction of $3.5^{4.0}_{-2.6}\cdot 10^{-8}$, and other calculations which
exploit the Heavy Quark Chiral Perturbation Theory lead to similar results
\cite{Burdman:2003rs, Fajfer:2001jq}. 

In various new physics scenarios this process could be enhanced and it has been
pointed out that within the MSSM scenario gluino exchange could enhance the SM
rate of more than 2 orders of magnitude~\cite{Prelovsek:2000xy}.

\subsection{Search for \dgammagamma at \babar}

The present best limit on the \dgammagamma branching fraction is due to an
analysis performed at \babar and reported in Ref.~\cite{Lees:2011qz}. 
This analysis is based on 470.5 \fb of $e^+ e^-$ collisions at 
$\sqrt{s} = 10.58$ GeV and 10.54 GeV. 
The search for \dgammagamma was done in parallel with a
measurement of the $D^0 \to \pi^0 \pi^0$ branching fraction which is also the
main background for the di-photon final state. As normalisation channel was
chosen the $D^0 \to K^0_S \pi^0$ decay for its well known branching fraction. 

MC simulations were used to optimize the search for \dgammagamma decays. 
\dgammagamma ($D^0 \to \pi^0 \pi^0$) candidates were formed with pairs of
photons (pions) required to have an invariant mass between 1.7 and 2.1 \gevcc
(1.65 and 2.05 \gevcc). Photons were required to have a CM energy between
0.74 and 4 GeV. The $D^0$ candidates were required to come from a \dstdpi decay
and therefore kinematically fitted, together with a $\pi^+$,  to a common vertex
constrained to be within the beam spot. 

Banckground from B decays was rejected with a 2.85 (2.4) GeV/c momentum
threshold on the $D^{\ast}$ candidate for \dgammagamma ($D^0 \to \pi^0 \pi^0$),
while the large QED background was removed with a cut on the tracks and neutrals
multiplicity of the events. 
Finally the $D^0 \to \pi^0 \pi^0$ background to \dgammagamma was rejected with
an ad-hoc $\pi^0$-veto which required the photons not to be consistent with
coming from a $\pi^0$ decay after being combined with other photons of
the same event. 

\begin{figure}
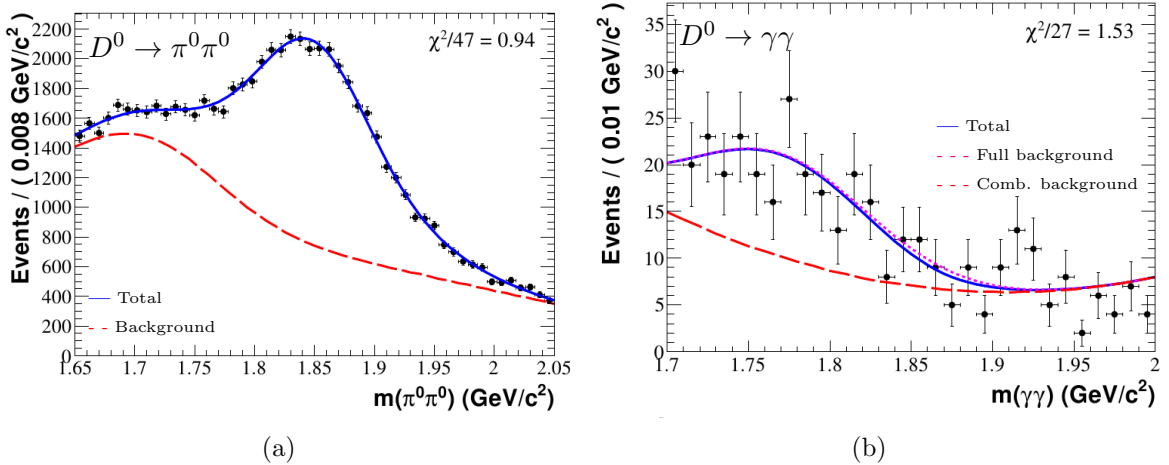

\subfigure[]{
 \begin{overpic}[width = 0.5 \textwidth ]{babar_d0pipi_mass.pdf}
 \put(15, 63){\small $D^0 \to \pi^0 \pi^0$}
 \put(15, 20){\tiny \textbf{\textcolor{blue}{\textemdash}} Total }
 \put(15, 15){\tiny \textcolor{red}{- -} Background }
\end{overpic}}
\subfigure[]{
\begin{overpic}[width =  0.5 \textwidth ]{babar_d0gg_mass.pdf}
\put(15, 65){\small $D^0 \to \gamma \gamma$}
 \put(60, 50){\tiny \textbf{\textcolor{blue}{\textemdash}} Total }
 \put(60, 45){\tiny \textcolor{magenta}{- - -} Full background }
 \put(60, 40){\tiny \textcolor{red}{- - -} Comb. background }
\end{overpic}}
\caption{Invariant mass distribution for (a) $D^0 \to \pi^0 \pi^0$ and (b)
\dgammagamma candidates in the analysis performed by the \babar collaboration
\cite{Lees:2011qz}. }\label{fig:d0gammagamma_babar}
\end{figure}

The invariant mass of $D^0 \to \pi^0 \pi^0$ candidates is shown in
Fig.~\ref{fig:d0gammagamma_babar}(a); the superimposed fitted PDF is composed
of a $3^{rd}$ order Chebychev polynomial for the background and of a signal PDF
which is the sum of a Crystal Ball, a gaussian and a bifurcated gaussian
function. A signal yield of about 26k events was estimated which led to a
branching fraction measurement of:  $\mathcal{B}(D^0 \to \pi^0 \pi^0) = (8.4 \pm
0.1 \pm 0.3)\times 10^{-4}$. 
On the other hand a negative signal yield, consistent with zero, was estimated
from the fit to the \dgammagamma invariant mass distribution, shown
in Fig.~\ref{fig:d0gammagamma_babar}(b).
The measured upper limit on the branching fraction of this decay was estimated
to be: $\mathcal{B}(\dgammagamma) < 2.2 \times 10^{-6}$ at 90\% confidence
level. 
While this result is not yet approaching the SM rate, it limits the
branching fraction to, at most, $\sim70$ times the SM levels, tightly
constraining New Physics models. 

\section{\dellell}
\label{sec:dellell}

While the two-body leptonic signature of the \dellell decays is more appealing
from the experimental point of view, theoretically they are foreseen to be even
more rare than \ctoull processes, due to the helicity suppression of the phase
space. 
The Standard Model short distance contribution for \dmumu is at the level of
$10^{-18}$ and even less for \dee. The long distance contributions are
relevant in the SM branching fraction and in particular the dominant one is due
to the two-photon intermediate state and the following expression  can be
derived for the di-muon final state~\cite{Burdman:2001tf}:
\begin{equation}
\label{eq:dmumu}
{\cal B}(D^0\to \mu^+\mu^-) \simeq  2.7 \times 10^{-5} {\cal B}
(D^0 \to \gamma
\gamma)
\end{equation}
which leads to an estimate of a minimum theoretical contribution to the 
branching fraction of  ${\cal B}(D^0\to \mu^+\mu^-) \gtrsim  10^{-13}$. However
equation \eqref{eq:dmumu} could also be exploited to estimate a rough 
upper limit to this contribution:  together with the experimental limit on
\dgammagamma set by \babar~\cite{Lees:2011qz} described earlier, this gives
\mbox{${\cal B}(D^0\to \mu^+\mu^-)_{\gamma\gamma} \lesssim 6 \times 10^{-11}$}. 

Many NP models can instead enhance the \dellell branching fraction by various
orders of magnitude. Moreover it has been noted that the \dmumu decay can be
related directly to the couplings involved in the $D^0 -\bar D^0$ mixing
\cite{Golowich:2009ii} so that limits on one of the two processes can be used
to constrain evaluations of the other. 
As an example one of the largest contributions is from $\displaystyle{\not}
R_p$ SUSY and gives an estimate of~\cite{Golowich:2009ii}: 
\begin{equation}
{\cal B}^{\not R_p}_{D^0 \to \mu^+\mu^-}  \simeq 4.8 \times 10^{-7}  x_{\rm D}
\left( {300~{\rm GeV}\over 
m_{\tilde{d}_k}} \right)^2 \ 
\le \ 4.8\times 10^{-9} \left( {300~{\rm GeV}\over 
m_{\tilde{d}_k}} \right)^2
\end{equation}
which is dependent on the mass of the down type quarks super partners. 

Finally, also contributions from Leptoquarks were found to be relevant, and
predicted branching fractions at the level of $5\times
10^{-7}$~\cite{Benbrik:2008ik}  and are being already strongly constrained by
experimental limits. 

In the following we will present a preliminary result of the LHCb experiment on
the \dmumu decay, which is the strongest limit up to date, and the results
obtained
by the Belle collaboration on all the \dellell decays. 

\subsection{Search for \dmumu at LHCb}
\label{sec:dmumu_lhcb}

The search for the \dmumu decay at LHCb is based on a dataset of 0.9~\fb of
$pp$ collisions in LHC at $\sqrt{s} = 7$ TeV \cite{LHCb-CONF-2012-005}. An
additional sample of about 79~\pb was also used in order to optimize
the selection but not used in the final analysis. 

\begin{figure}
\subfigure[]{\begin{overpic}[width = 0.5 \textwidth]{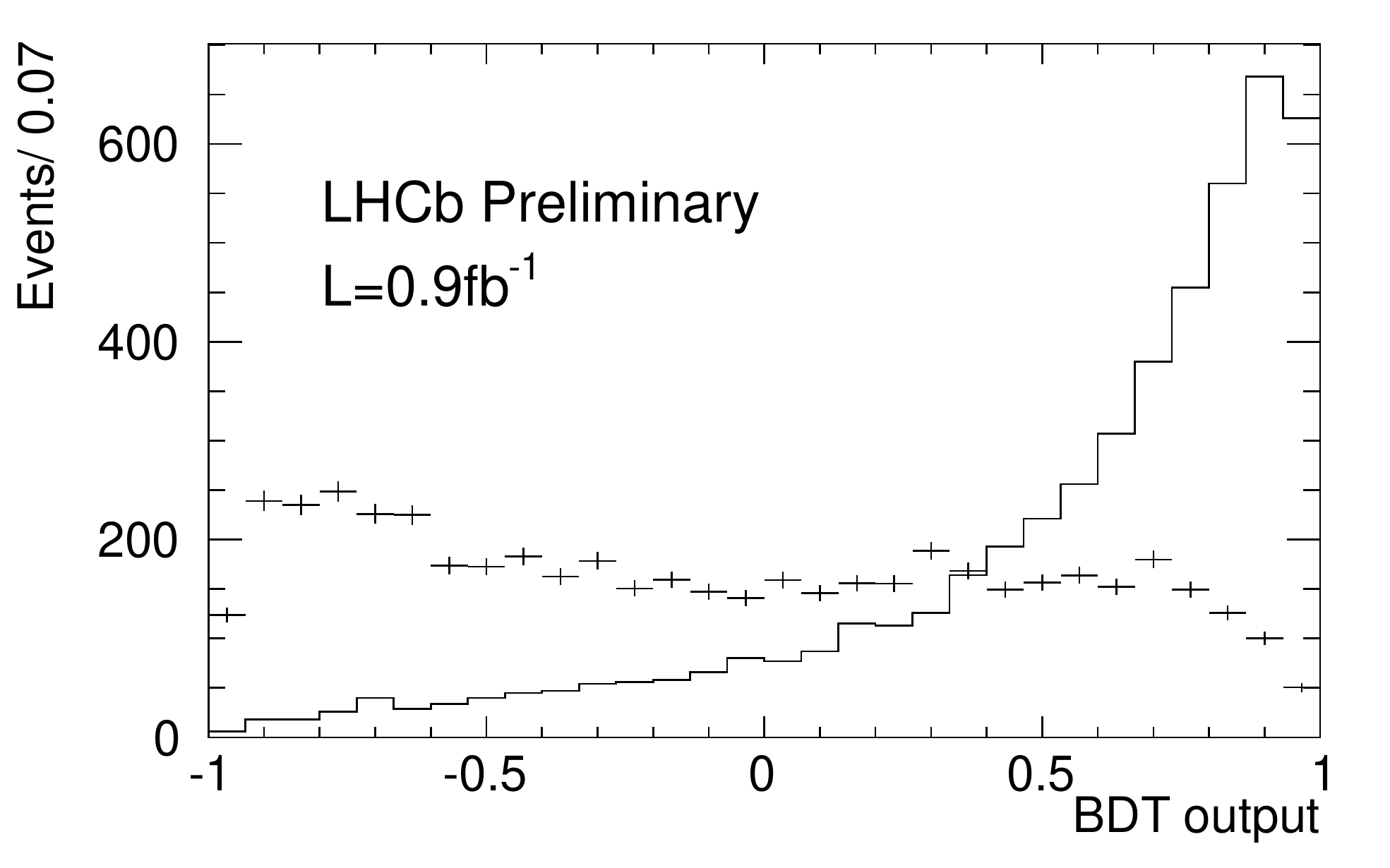}
 \put(25, 35){{\tiny \textemdash \quad \dmumu signal }}
 \put(25, 30){{\tiny + \quad Comb. background }}
\end{overpic}}
\subfigure[]{\begin{overpic}[width = 0.5 \textwidth 
]{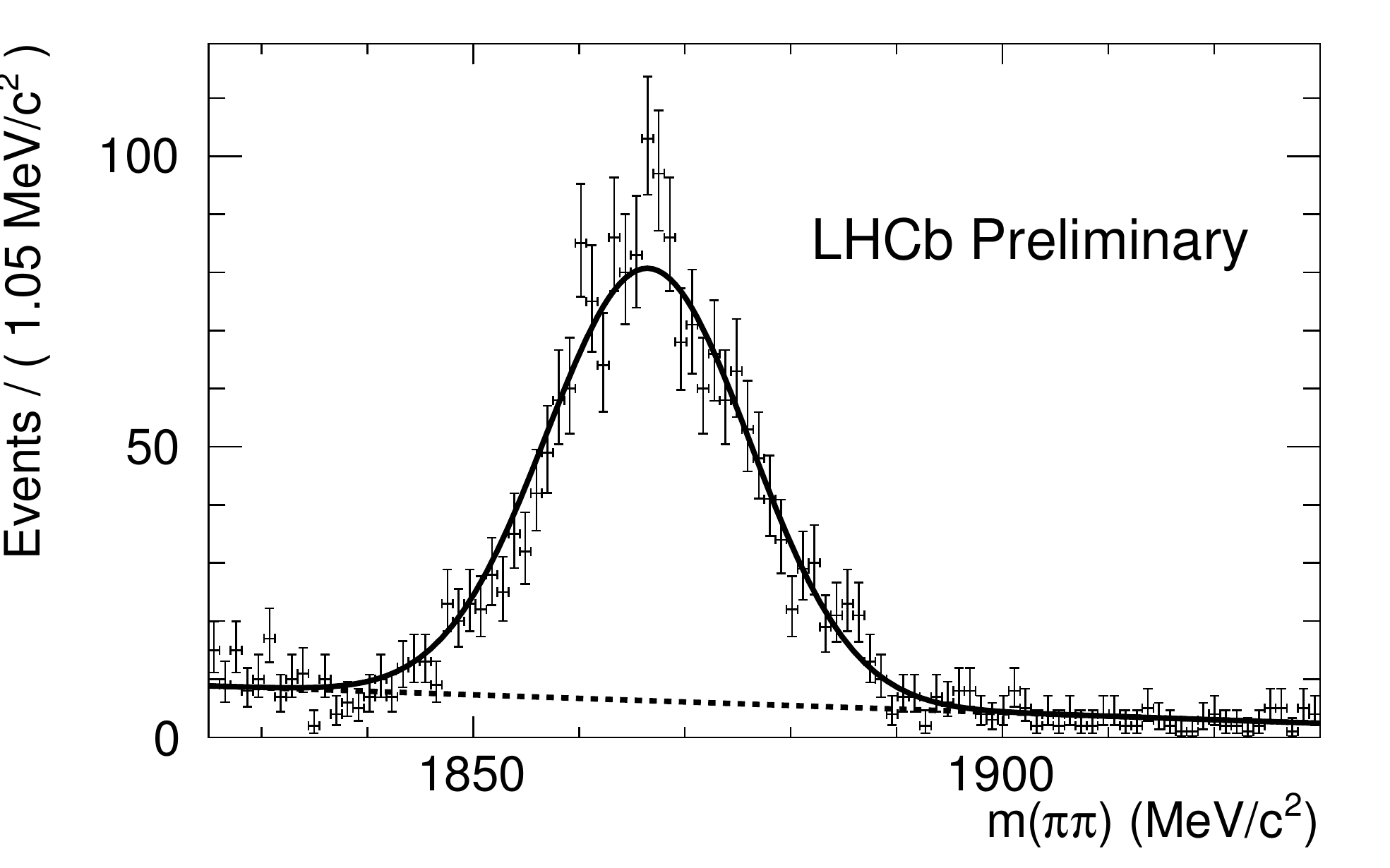}
\put(60,35){$D^0 \to \pi^+ \pi^-$}
\end{overpic}}
\caption{In (a) the distribution of the output of the BDT for signal MC
simulations (continuous line) and combinatorial background from data
(crosses) are shown as obtained in the LHCb experiment. In (b) is shown the
invariant mass distribution for \dpipi candidates in data
\cite{LHCb-CONF-2012-005}.}\label{fig:d0mumu_1}
\end{figure}

\begin{figure}
\subfigure[]{
\begin{overpic}[width = 0.5 \textwidth ]{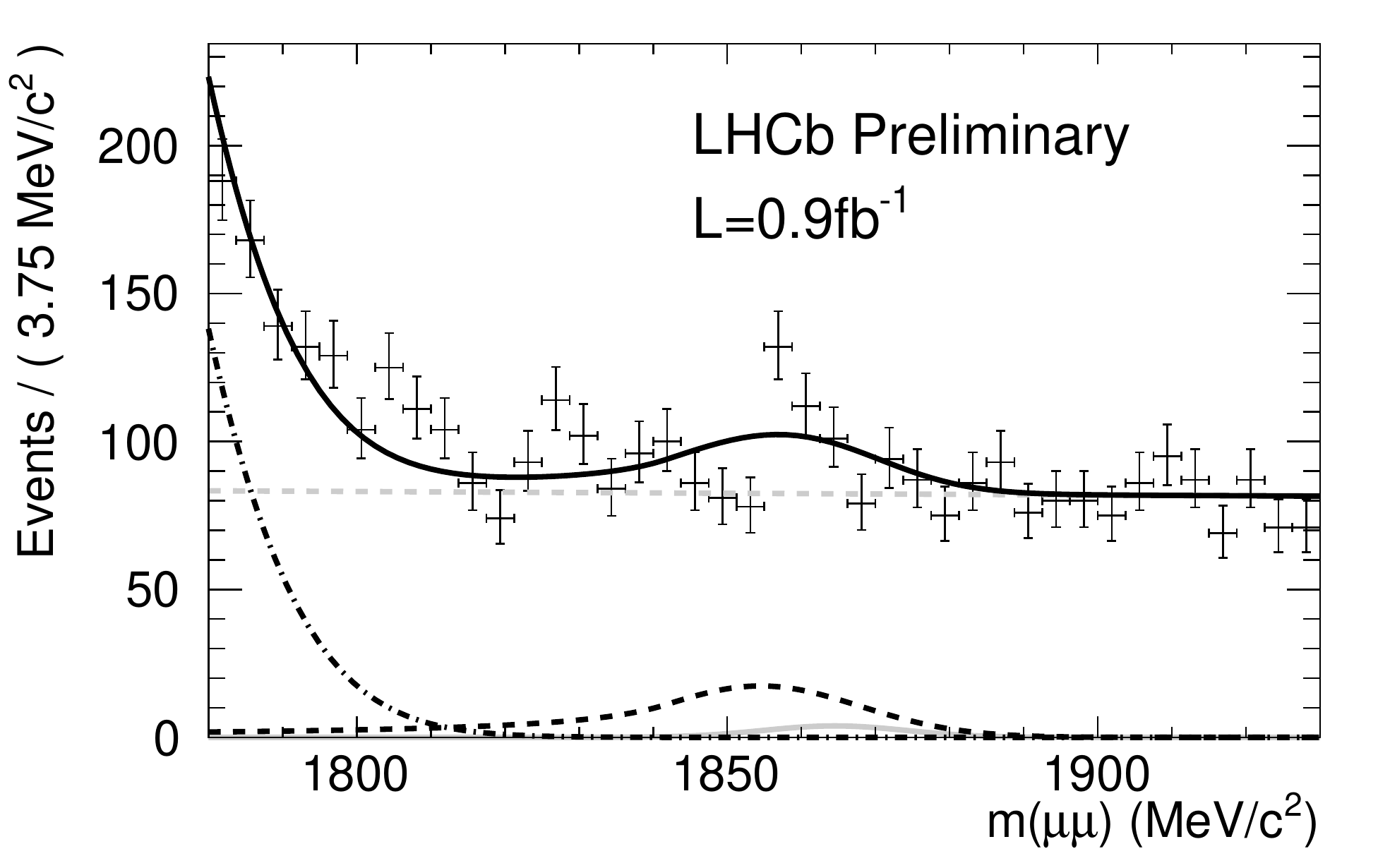}
  \put(50,40){\tiny $142<\Delta M<149$}
  \put(18,50){\dmumu}
\end{overpic}}
\subfigure[]{\includegraphics[width = 0.5 \textwidth ]{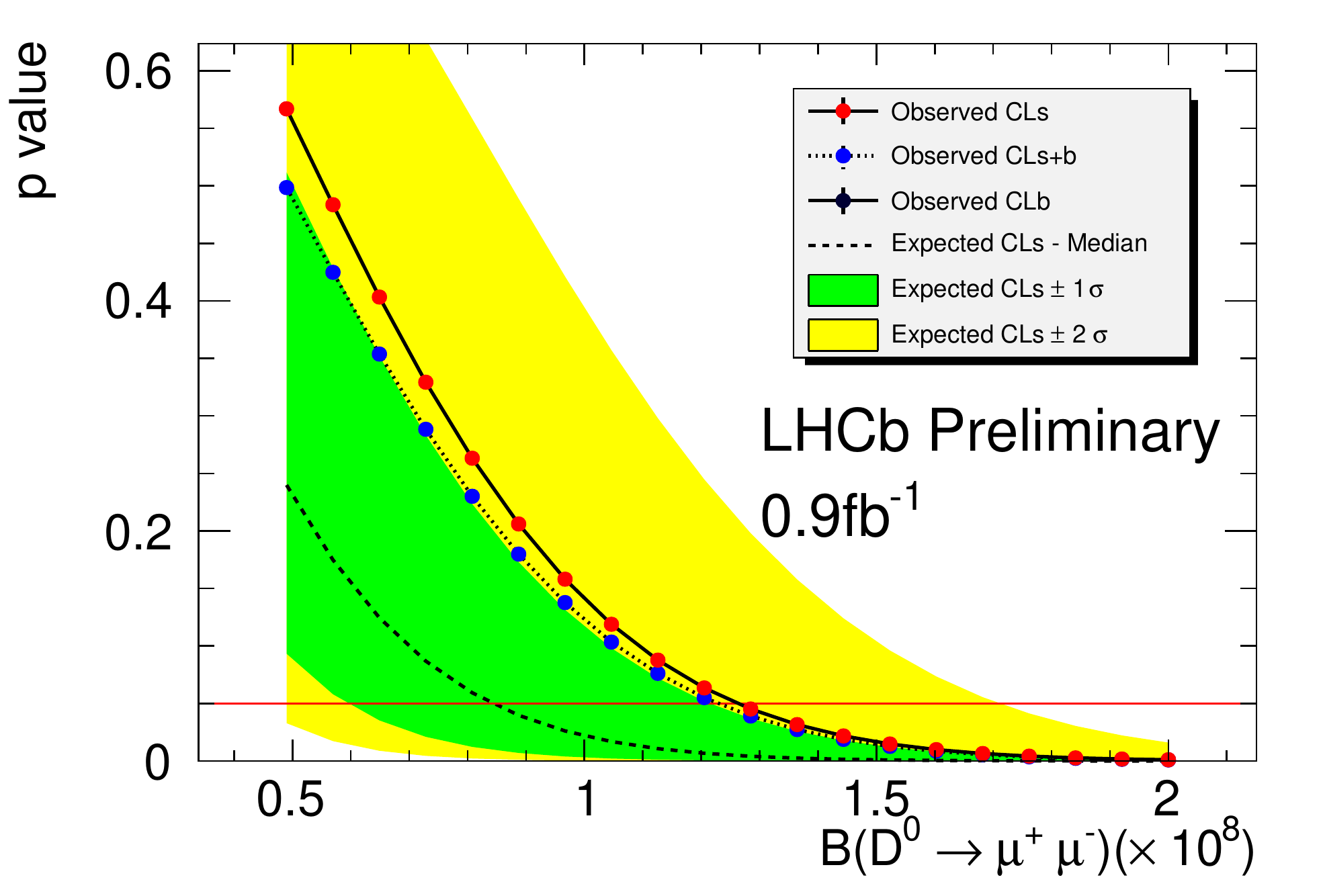}
}
\caption{(a) Invariant mass distribution for \dmumu
candidates, projected from the $M_D, \Delta M$ plane, and
the fit to the distribution: the black line is the total
fit function composed of the combinatorial background (light grey dashed line),
\dpipi mis-identified background (dark dashed line), \dkpi mis-identified
background (dash-dotted line) and the signal \dmumu (light grey continuous
line).   In (b) the CLs value as a function of the \dmumu branching
fraction as obtained with the CLs method is shown; the red line shows the upper
limit to the branching fraction at 95 \%
CL~\cite{LHCb-CONF-2012-005}.}\label{fig:d0mumu_2}
\end{figure}

Pairs of opposite side muons were selected to form $D^0$ candidates, which were
required to come from a \dstdpi decay. After a first selection based on
standard geometrical and kinematic variables, a Boosted Decision Tree (BDT) was
built in order to reject the combinatorial background, combining information
from the following variables: $\chi^2$ of the impact parameter of the $D^0$
candidate and pointing angle of the $D^0$ with respect to the primary vertex,
minimum transverse momentum of the
two muons, minimum $\chi^2$ of impact parameter of the two muons, angle of
the positive muon in the $D^0$ candidate rest frame with respect to the $D^0$
flight direction. In Fig.~\ref{fig:d0mumu_1} is shown the distribution of the
BDT output for signal MC events and for events in the di-muon invariant mass
sidebands. The final cut value on the BDT was chosen in order to have the best
expected upper limit on the \dmumu branching fraction. 
A large background to this decay is the \dpipi with both pions mis-identified
into muons. This background was estimated by measuring the mis-ID probability
directly on data \dkpi decays. The double mis-ID probability was estimated to
be  $p(\dpipi\to \mu \mu) = (27.3\pm3.4\pm2.0)\cdot10^{-6}$. 
The \dstdpipi decay was also exploited as normalisation channel, after being
selected with the same selection as the signal, with the exception of the muon
identification. The \dpipi yield was estimated with a 2-dimensional fit to the
distribution in the $\Delta M$ versus
invariant mass plane. A projection of this distribution together with the fit
is shown in Fig.~\ref{fig:d0mumu_1}(b). 
The signal and normalisation channel efficiencies were estimated from MC
simulations and corrected for data-MC discrepancies. 
The estimate of the \dmumu signal yield was also done in the 2D $\Delta
M$-$M_{D^0}$ plane. The data distribution was fitted with a PDF built upon the
following components: the signal was parametrised as a gaussian (in M) times a 
double gaussian (\dm), the \dpipi mis-ID as Crystal Ball (M) times a
double gaussian (\dm), the  \dkpi tail mis-ID as a gaussian in both M and \dm
and the combinatorial background as an exponential (in M) times the following
function: $ f(\dm) = \left(\frac{\dm}{a}\right)^2 ( 1 - e^{\frac{\dm -
\dm_0}{c}}) + b \cdot (\frac{\dm}{d} - 1 )$. 
The projection of this fit along the $D^0$ candidate invariant mass is shown in
Fig.~\ref{fig:d0mumu_2}(a). As no significant signal was observed, the upper
limit to the \dmumu branching fraction was estimated with the CLs method to be: 
$\mathcal{B}(\dmumu) < 1.3 (1.1) \times 10^{-8}$ at 95 (90)\% CL. 
In Fig.~\ref{fig:d0mumu_2}(b) is shown the CLs value as a function of the
\dmumu branching fraction and the expected values and bands. 
This limit is still several orders of magnitude far from the SM predictions
but constraints various NP models. 

\subsection{Search for \dellell at Belle }
\label{sec:dll_belle}

\begin{figure}
\begin{center}
\includegraphics[width = 8 cm]{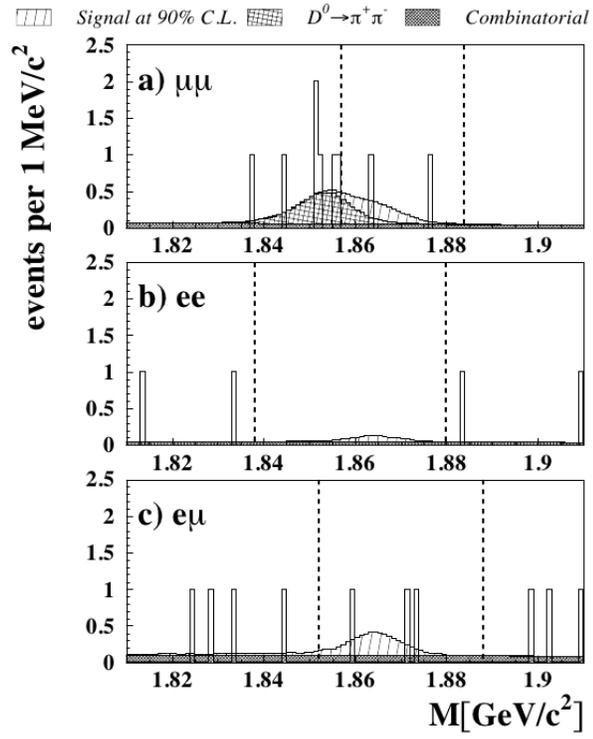}                          
                               \end{center}
\caption{Invariant mass distributions for \dellell
candidates in the analysis performed by the Belle
Collaboration \cite{Petric:2010yt}.}\label{fig:d0ll_belle}
\end{figure}

Another search for di-leptonic $D^0$ decays was done by the Belle collaboration
and reported in Ref.~\cite{Petric:2010yt}. 
This analysis was based on 660 \fb of $e^+e^-$ collisions at the $\Upsilon(4S)$
and just below it, collected with the Belle detector at KEKB. 
$D^0$ candidates were built from two opposite side leptons in the \dmumu, \dee
and \demu final states. Only $D$ mesons from $e^+e^- \to c\bar c$ were used,
excluding those from $B$ decays because of the higher combinatorial
background; these were removed by requiring a maximum allowed missing energy
which is high in events with B semileptonic decays due to missing neutrinos. 
The \dpipi final state was also reconstructed as normalisation and control
channel. Standard particle identification criteria, based on $dE/dx$, TOF,
Cherenkov light and muon detectors information, were exploited in order to
select pions, electrons and muons. 
The $D^0$ candidate was required to come from a \dstdpi decay from the vertex
of the $e^+ e^-$ interaction point. Further rejection of combinatorial
background and of B-decays was achieved by requiring a $D^{\ast+}$ candidate
momentum in excess of 2.5 GeV/c. Candidates were selected to be in a region of
the $D^0$ candidate invariant mass between 1.81 and 1.91 \gevcc and with
$q<20$ MeV, where $q = (M_{D^{\ast+}} - M_{D^0} - m_{\pi})c^2$. 
Further selection criteria, based on the missing energy, lepton
identification and signal region in the mass versus $q$ plane, were optimized
differently for each final state with a blind analysis. 
Residual combinatorial background was estimated from the sideband region of the
$M-q$ plane, while the mis-identified \dpipi background was evaluated from the
decay itself reconstructed in data, re-weighted for momentum dependent
mis-identification probabilities. 
The invariant mass distributions after the final selection criteria are shown
in Fig.~\ref{fig:d0ll_belle}. No events in excess of the estimated background
are observed in any of the final states. 
A binned maximum likelihood fit to the invariant mass distribution was used to
estimate the yield of the normalisation channel \dpipi. Efficiencies were
estimated from tuned MC samples. Finally the following upper limits
were derived on the signal channels branching fractions: $\mathcal{B}(\dmumu) <
1.4 \times 10^{-7}$, $\mathcal{B}(\dee) <
7.9 \times 10^{-7}$, $\mathcal{B}(\demu) <
2.6 \times 10^{-7}$ at 90\% CL. 

The ones on \dee and \demu are the best limits on these channels to date 
and limit further the parameter space for NP models. 

\section{Forbidden decays}
\label{sec:forbidden}

While rare decays are often dominated by the theoretical uncertainties, 
an unambiguous sign of New Physics would be to observe one of the decays which
in the Standard Model are forbidden because of violation of one or more
conservation laws. In particular here we consider decays which do not conserve
the Lepton Family Number, the total Lepton Number, the Baryon Number or more
than one of these.

\begin{figure}
\begin{overpic}[width = \textwidth ]{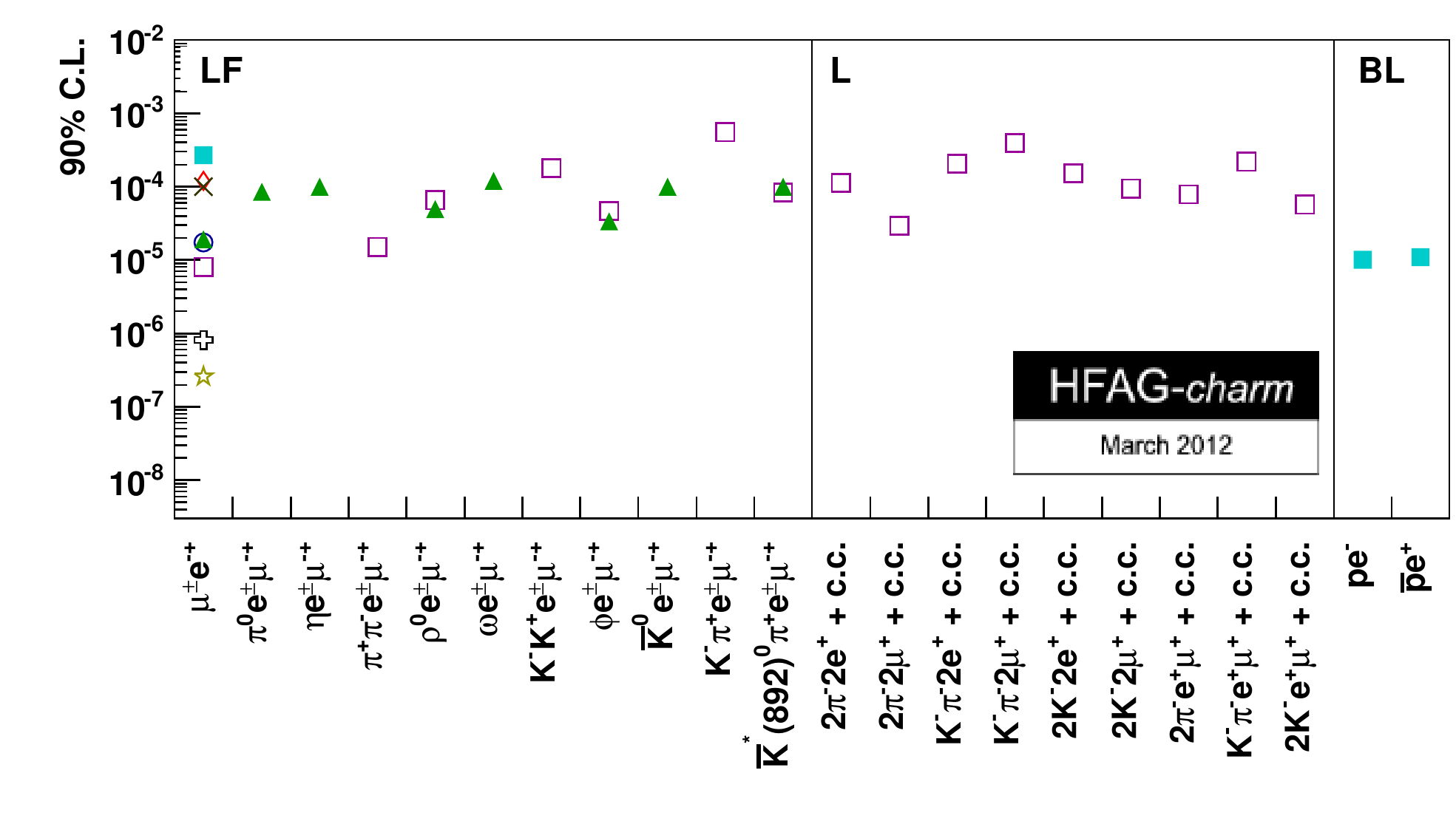}
 \put(35, 22){\includegraphics[width = 2.5 cm ]{D0_leg.pdf}}
\end{overpic}
\caption{Summary of branching fraction upper limits on
$D^0$ forbidden decays, compiled by the HFAG group
\cite{Amhis:2012bh}. }\label{fig:lfv_hfag}
\end{figure}

Experimentally none of these decays has ever been observed, and therefore
limits on the branching fractions are set. A summary of the current situation
for
the neutral sector, \emph{i.e.} for $D^0$ decays, can be seen in
Fig~\ref{fig:lfv_hfag} where is shown a collection of the various limits on
branching fractions compiled by the HFAG group \cite{Amhis:2012bh}.
These upper limits have been set, with few exceptions, by the already mentioned
experiments: E791 \cite{Aitala:2000kk} and CLEOII \cite{Freyberger:1996it}.

\begin{figure}
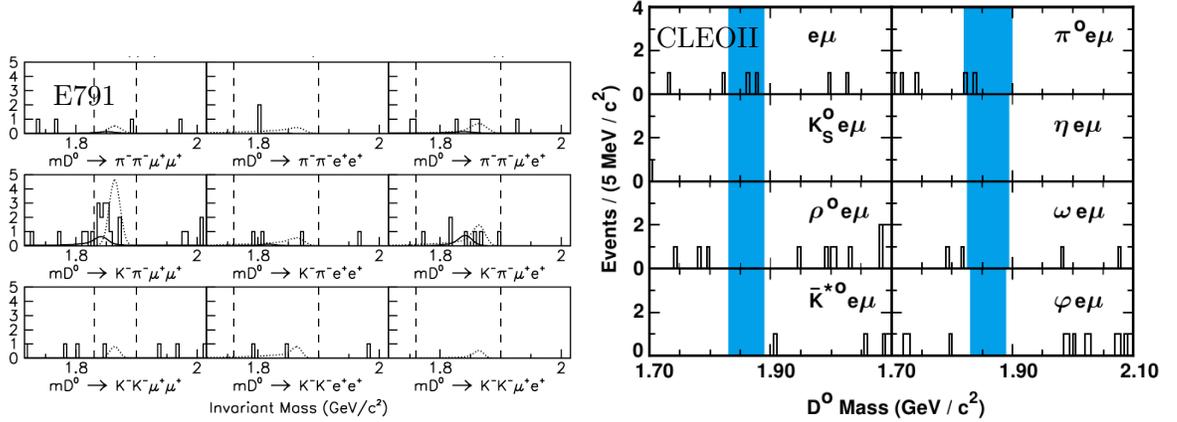

\subfigure{\begin{overpic}[width = 0.5 \textwidth]{E791_d0hmumu_LFV.pdf}
\put(8,55){\small E791}
\end{overpic}}
\subfigure{\begin{overpic}[width =  0.5
\textwidth]{CLEOII_results_LFV.pdf}
 \put(12,65){\small CLEOII}
\end{overpic}}
 \caption{Invariant mass distributions for candidates of $D^0 \to X
\ell \ell$ LFV decays from the E791~\cite{Aitala:2000kk} and CLEOII
\cite{Freyberger:1996it} experiments. }\label{fig:e791_cleo_lfv}
\end{figure}

Analyses of $D^0 \to X \ell \ell$ LFV decays were performed in the very same
way as described in \S\ref{sec:neutral_fcnc_e791} and
\S\ref{sec:neutral_fcnc_cleoII} for the E791 and CLEOII experiments
respectively. The invariant mass distributions for some of the analysed final
states are shown in Fig.~\ref{fig:e791_cleo_lfv}. None of the searched decays
was observed and the limits on the branching fractions are between $10^{-5}$ and
few in $10^{-4}$. 

One of the exceptions in this sector is the already mentioned limit on \demu
which has been set by the Belle collaboration \cite{Petric:2010yt} and described
in \S\ref{sec:dll_belle}.

\begin{figure}
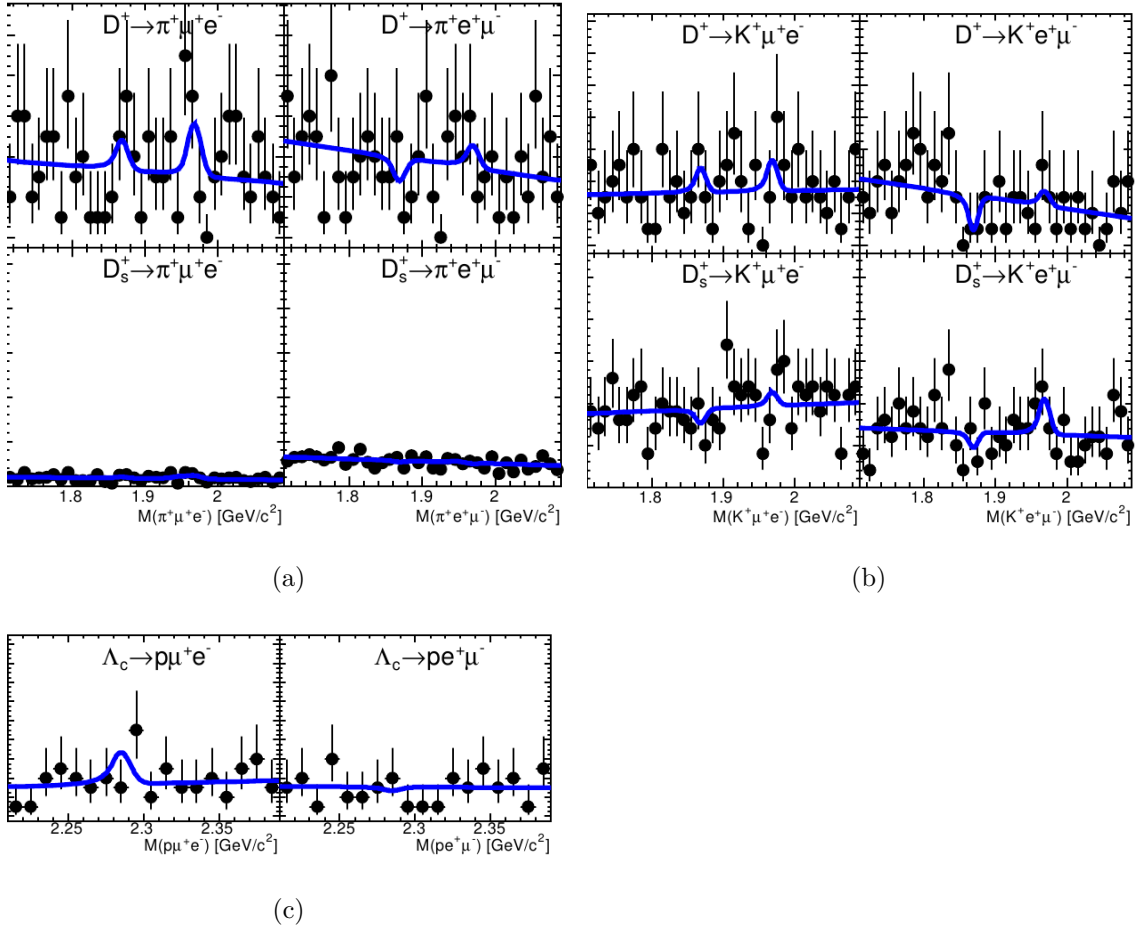

  \subfigure[]{\includegraphics[width = 0.5
\textwidth]{babar_Dpill_LFV.pdf}}
 \subfigure[]{\includegraphics[width = 0.5
\textwidth]{babar_DKll_LFV.pdf}}
\subfigure[]{\includegraphics[width = 0.5
\textwidth]{babar_Lcpll_LFV.pdf}}
\caption{Invariant mass distributions of charged charm
hadron decays into SM forbidden final states \cite{Lees:2011qz}.
}\label{fig:babar_lfv}
\end{figure}

The \babar collaboration has instead the
almost absolute monopoly on the limits on the charged sector as can be seen by
the HFAG summary plots in Fig.~\ref{fig:hfag_charged}. Most of the
limits on LFV and BV decays for $D^+$,$D^{+}_s$ and
$\Lambda^+_c$ come in fact from the already discussed Ref.~\cite{Lees:2011qz}
(Cfr.\S\ref{sec:charged_fcnc_babar}).  The analysis of forbidden \xplusctohll
final states was the same as for the FCNC ones. Invariant mass distributions of
some of the analysed states are shown in Fig.~\ref{fig:babar_lfv} where it can
be seen that no signal was observed. The upper limits on the branching
fractions are all of the order of $10^{-5}$.

\section{Conclusion}

We have given a broad overview of the current experimental status
of searches for charm rare decays. While in the past less interest was
devoted to these decays than to the $B$ and $K$ correspondent ones, a renewed
effort is going on both on the theoretical and experimental side. This is
partially motivated by the recent measurement in the other charm processes but
also due to new experimental searches going on at LHC and B-factories. 

The most stringent limits are still not approaching the Standard Model
predictions and large theoretical errors still dominate the calculations of SM
long distance contributions. Nevertheless these searches are already
constraining the parameter space of various New Physics models. 

Furthermore, as we have seen,  there is a large playground of searches,
especially in the neutral charm sector, which are more than ten years old and
could be repeated and updated in current experiments, improving the knowledge
of this interesting field.

\bibliographystyle{iopart-num}
 \bibliography{Bibliography}

\end{document}